\documentclass[epj]{webofc}
\usepackage[utf8]{inputenc}
\usepackage[varg]{txfonts}   
\usepackage{booktabs}
\usepackage{xcolor}
\definecolor{darkred}{rgb}{0.4,0.0,0.0}
\definecolor{darkgreen}{rgb}{0.0,0.4,0.0}
\definecolor{darkblue}{rgb}{0.0,0.0,0.4}
\usepackage[bookmarks,linktocpage,colorlinks,
    linkcolor = darkred,
    urlcolor  = darkblue,
    citecolor = darkgreen]{hyperref}
\usepackage{subfigure}
\usepackage{wrapfig}
\wocname{EPJ Web of Conferences}
\woctitle{Lattice2017}
\begin{document}
%
\selectlanguage{english}
\title{%
Improved real-time dynamics from imaginary frequency lattice simulations
}
\author{%
\firstname{Jan~M.} \lastname{Pawlowski}\inst{1,2}\fnsep\and
\firstname{Alexander} \lastname{Rothkopf}\inst{3}\fnsep\thanks{Speaker, \email{rothkopf@thphys.uni-heidelberg.de}}
}
\institute{%
Institut f\"{u}r Theoretische Physik, Universit\"{a}t Heidelberg, Philosophenweg 16, 69120 Heidelberg, Germany
\and
ExtreMe Matter Institute EMMI, GSI Helmholtzzentrum f\"{u}r Schwerionenforschung mbH, 64291 Darmstadt, Germany
\and
Institut f\"{u}r Theoretische Physik, Universit\"{a}t Heidelberg, Philosophenweg 12, 69120 Heidelberg, Germany
}
\abstract{%
The computation of real-time properties, such as transport coefficients or bound state spectra of strongly interacting quantum fields in thermal equilibrium is a pressing matter. Since the sign problem prevents a direct evaluation of these quantities, lattice data needs to be analytically continued from the Euclidean domain of the simulation to Minkowski time, in general an ill-posed inverse problem. Here we report on a novel approach to improve the determination of real-time information in the form of spectral functions by setting up a simulation prescription in imaginary frequencies. By carefully distinguishing between initial conditions and quantum dynamics one obtains access to correlation functions also outside the conventional Matsubara frequencies. In particular the range between $\omega_0$ and $\omega_1=2\pi T$, which is most relevant for the inverse problem may be more highly resolved. In combination with the fact that in imaginary frequencies the kernel of the inverse problem is not an exponential but only a rational function we observe significant improvements in the reconstruction of spectral functions, demonstrated in a simple 0+1 dimensional scalar field theory toy model.}
\maketitle
\section{Motivation}\label{intro}

Lattice QCD simulations at finite temperature have seen dramatic progress over the past decade and by now are capable of computing quantities of vital phenomenological interest, such as the QCD equation of state with dynamical u,d, and s quarks in the continuum limit \cite{Borsanyi:2016ksw,Bazavov:2014pvz,Borsanyi:2013bia} with nearly physical pion masses (even venturing towards finite Baryo-chemical potential \cite{Bazavov:2017dus,Gunther:2016vcp}). Two major collaborations explore these static observables independently, based on different fermion actions with compatible outcomes, i.e. the results are indeed highly robust.

The situation is very different when it comes to the determination of dynamic properties of QCD matter, where results are still scarce and have not yet reached a similar accuracy and robustness. One prominent example are QCD transport coefficients, such as the shear \cite{Pasztor:2016wxq,Mages:2015rea,Meyer:2009jp,Meyer:2007ic} and bulk \cite{Meyer:2007dy} viscosity, which are central ingredients in the hydrodynamic description of relativistic heavy-ion collisions at RHIC and LHC \cite{Bernhard:2016tnd}. Another example are in-medium meson spectral functions (see e.g. \cite{Kim:2014iga,Aarts:2014cda}) that are vital to the understanding of heavy-quarkonium bound states in heavy-ion collider experiments.

The underlying difficulties are related to the fact that lattice computations are carried out in an artificial Euclidean time. This in turn necessitates an analytic continuation of simulated correlation functions, before genuine real-time information can be extracted. It is in principle possible to carry out this task, since both Euclidean and Minkowski correlators are described by the same real-valued spectral function, to which they are connected via integral relations. In turn the real-time quantities of interest may often be expressed directly in term of these spectral functions, in case of the shear viscosity $\eta=-\frac{1}{\pi} \lim_{\mu\to0^+}\lim_{p\to0^+} \frac{\rho^{\pi\pi}(\mu,p)}{\mu}$
with $\rho^{\pi\pi}(\mu)$ denoting the spectral function of the spatial part of the traceless energy momentum tensor correlator in real-time frequencies $\mu$. In practice the task of extracting spectral functions is highly ill-posed, as a continuous functions needs to be unfolded from a noisy and finite set of simulated data. Bayesian inference has been found to provide a highly efficient tool for carrying out such unfolding tasks and in this study we will deploy the recently developed BR method \cite{Burnier:2013nla} whenever spectra are reconstructed.The standard relations for bosonic fields read
\begin{align}
G(\tau)=\int_0^\infty \,d\mu \, \frac{{\rm cosh}[\mu(\tau-\beta/2)]}{{\rm sinh}[\mu \beta/2]}\,\rho(\mu), \quad \overset{{\rm Fourier }}{\longleftrightarrow} \quad  G(i\omega_n)=\int_0^\infty \,d\mu\,\frac{2\mu}{\mu^2+\omega_n^2}\,\rho(\mu).\label{eq:CorrSpecRel}
\end{align}
Two concrete challenges manifest themselves in the above equations. First, the inverse problem from Euclidean data will suffer from exponential information loss induced by the strongly damping cosh/sinh kernel. This issue was well illustrated in \cite{Pasztor:2016wxq}, in that the Euclidean correlator is only marginally sensitive to sizable changes in the spectral function, especially at small frequencies. One possible step towards resolving this problem would be to go to imaginary frequencies, where the K\"all\'en-Lehmann kernel is only a rational function and thus the correlator exhibits much more sensitivity to spectral features, also at small $\mu$.

This however is only a partial solution since another more fundamental issue persists. In the standard Euclidean formulation of thermal field theory, the imaginary time axis is compactified and its physical extent related to the inverse temperature of the system. Thus the imaginary frequency correlator is resolved only on the Matsubara frequencies $\omega_n=2\pi T n$, even though the right expression of eq.\eqref{eq:CorrSpecRel} indicates that $G(i\omega)$ exists also at intermediate frequencies. The physics of transport coefficients is indeed most prominently encoded in $G(i\omega)$ between $\omega_0=0$ and $\omega_1=2\pi T$ and already at $\omega_2$ only contributes marginally to the correlator. Resolving the Euclidean domain more and more finely, i.e. going to the continuum limit, only gives access to higher Matsubara frequencies but does not resolve the regime relevant in practice any better. I.e. from the point of view of the inverse problem it would be most advantageous to resolve this low imaginary frequency regime, which then significantly improves the chance of a successful reconstruction. Of course if all Matsubara frequencies were known exactly, analyticity permits us to extract the full real-time information already from the discrete Matsubara information. In practice however both the number and precision of the data is limited. In order to make progress in the determination of transport properties, we have recently devised a novel approach to thermal quantum fields \cite{Pawlowski:2016eck}, whose ability to dramatically improve the reconstruction of spectral functions is demonstrated in a scalar toy-model in (0+1) dimensions.
\section{The non-compact Euclidean time setup}\label{sec-1}

Starting point for the novel approach is to formulate the thermal field theory as genuine initial value problem on the Schwinger-Keldysh contour, with a forward propagating branch populated by fields $\varphi^+$(t), a backward propagating branch $\varphi^-$(t) and a compact Euclidean domain of $\varphi_E(\bar{\tau})$, which encodes the thermal initial conditions. The corresponding partition function reads
\begin{align}
Z=\int {\cal D}\varphi_Ee^{-S[\varphi_E]}\int_{\varphi^+(0)=\varphi_E(0)}^{\varphi^-(0)=\varphi_E(\beta)} {\cal D} [\varphi] e^{iS[\varphi^+]}\,e^{-iS[\varphi^-]},
\end{align}
where implicitly the fields at the latest real-time step $t_f$ are identified $\varphi^+(t_f)=\varphi^-(t_f)$. Note that here the Euclidean time $\bar{\tau}$ is nothing but a mathematical tool introduced to efficiently sample the thermal initial conditions for $\varphi^+$ and $\varphi^-$ provided by the density matrix $\rho={\rm exp}[-\beta H]$. \begin{wrapfigure}{r}{0.45\textwidth}
  \begin{center}\vspace{-0.65cm}
  \hspace{-0.5cm}\includegraphics[scale=0.53,trim=0.42cm 0.3cm 0 0.32cm, clip=true]{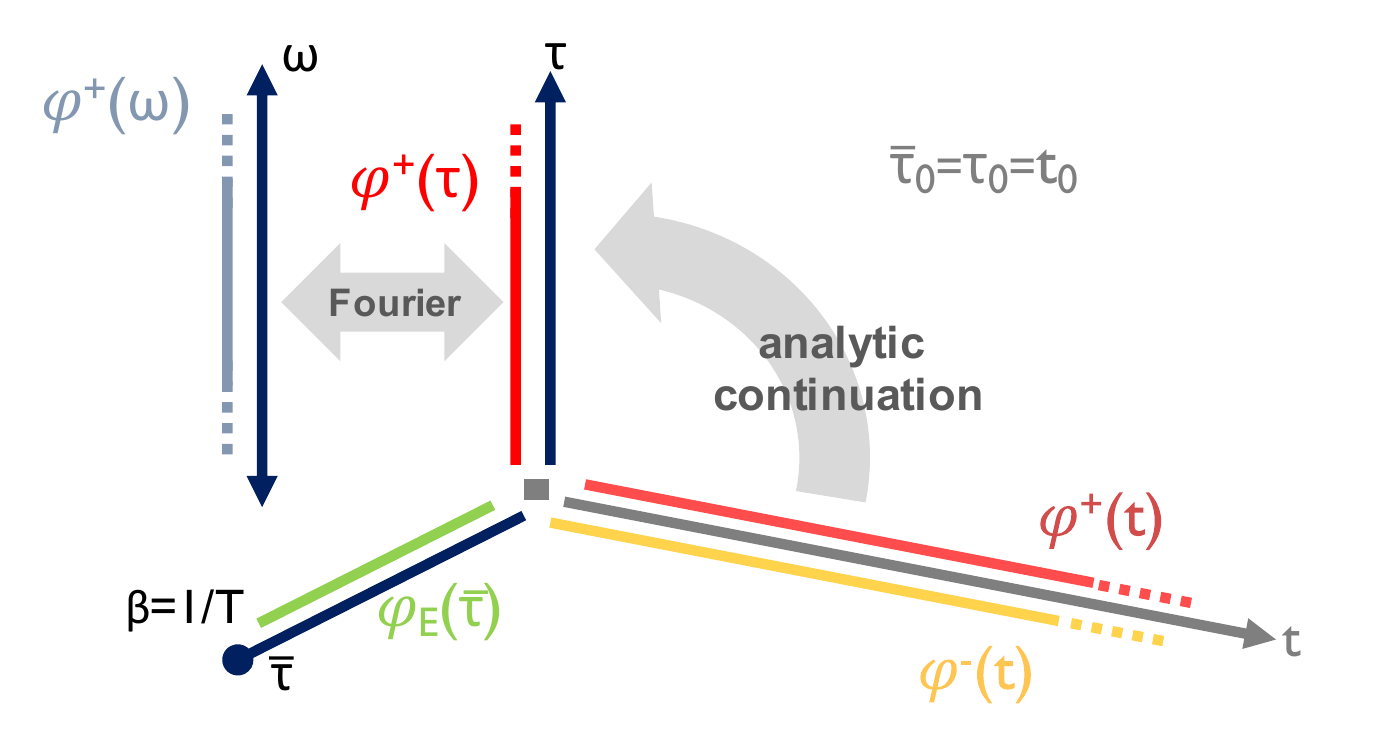}
  \end{center}\vspace{-0.5cm}
 \caption{Sketch of our setup, where, after effectively cutting open the real-time paths at infinity, we analytically continue the branches to an additional non-compact imaginary time domain $\tau$.}\label{Fig:Setup}\vspace{-1.75cm}
\end{wrapfigure}
What makes thermal equilibrium special is the fact that the spectral function of the system can be computed directly from the fields $\varphi^+$ in form of the correlator $G^{++}(t)=\langle \varphi^+(t)\varphi^+(0)\rangle$, with the following momentum space decomposition before 
\begin{align}
\nonumber&G^{++}(p^0,{\mathbf p}) = \int \frac{dq^0}{2\pi i} \frac{\rho(q^0,{\mathbf p})}{p^0-q^0+i\epsilon}-n_B(p^0)\rho(p^0,{\mathbf p})
\end{align}
and after analytic continuation to imaginary frequencies
\begin{align}
&G^{++}(ip^0,{\mathbf p})= \int \frac{dq^0}{2\pi} \frac{\rho(q^0,{\mathbf p})}{ip^0-q^0}.
\end{align} 
I.e. in contrast to non-equilibrium settings no cross correlations between $\varphi^+$ and $\varphi^-$ are needed. In addition due to time translational invariance we may move the initial time point $t_0=\bar{\tau}_0$ to negative infinity and extend the contour to positive infinity. 

At this stage we argue that, due to causality, no field $\varphi^+$ at finite time $t$ may know about the implicit identification of $\varphi^+$ and $\varphi^-$, now residing at $t\to\infty$, since all correlations damp away. In turn we consider the two real-time paths {\it effectively cut open} as indicated by the dots on the right in Fig.\ref{Fig:Setup}. The second step is to also change the way how the initial conditions are handled. Since the physics is time translational invariant, we can measure any real-time correlation functions of interest by inserting operators on the forward contour close to positive infinity, which again due to causality cannot influence the behavior at the starting point. Therefore we treat the Euclidean contour of $\varphi_E(\bar{\tau})$ as {\it genuinely compact} and identify $\varphi_E(0)=\varphi_E(\beta)$, which for a finite extent of the real-time path would of course not be permissible.

The final step consists of analytically continuing the time variable $t$ of $\varphi^+$ and $\varphi^-$ onto a new infinitely extended imaginary time axis. As we regard the Euclidean time $\bar{\tau}$ as a mathematical tool only, we do still have the freedom to continue to a new $\tau=it$. Due to the decoupling of the two fields at infinity, from now on we consider only the $\varphi^+$ and $\varphi_E$ degrees of freedom. With the extent of the imaginary time $\tau$ not being constrained, we may actually carry out a genuine Fourier transform and consider the system directly in imaginary frequencies, which now are not restricted to the conventional Matsubara ones. 

Since several unconventional steps were taken in constructing this approach, let us have a look at the outcome in a non-trivial low-dimensional scalar toy-model, i.e. $\varphi^4$ theory in (0+1) dimensions. This theory is equivalent to the quantum mechanical anharmonic oscillator and it may be solved by computing an appropriate Schr\"odinger equation, in turn providing us with independent data for correlation- and spectral functions. 

The computation is implemented via stochastic quantization on two finite grids with the same spacing $\Delta\bar\tau$, concurrently updating both the fields $\varphi_E$, encoding temperature, as well as the fields $\varphi^+$, from which the correlation function and spectrum is eventually determined. Based on the Euclidean action $S_E=S_E^0+S_E^{\rm int}=\int d\tau^\prime \, \left( \frac{1}{2} (\partial_{\tau^\prime} \varphi)^2+ \frac{1}{2} m^2\varphi^2+\frac{\lambda}{4!}\varphi^4\right)$ we set up  real-valued Langevin updates for both the compact and non-compact time domains, denote Monte-Carlo time as s and use independent Gaussian noises $\eta$ and $\eta^\prime$
\begin{align}
\partial_s \varphi_E(\bar{\tau},s)=-\frac{\delta S_E[\varphi_E]}{\delta \varphi_E}+\eta(\bar{\tau},s), \quad \partial_s \varphi^+(\tau,s)=-\frac{\delta S_E[\varphi^+]}{\delta \varphi^+}+\eta(\tau,s). \end{align}
Note that of course in a finite box we are not able to represent an infinitely extended imaginary time axis. I.e. we will incur a finite volume artifact from having to specify an artificial boundary condition at the final $\tau$. This issue already affects the free theory, due to the  presence of time derivatives in $S_E$. Putting a boundary by hand (periodic, Dirichlet, etc.) on a finite imaginary time path for $\varphi^+$ yields incorrect values for $G^{++}$. They however can be made to converge to the correct result by extending the $\tau$ domain further and further, as we have explicitly tested. I.e. even by simulating directly in the new imaginary time one can systematically approach the correct result, paying as a price the numerical costs related to a large imaginary time extent. 

The convergence to the infinite volume result may be significantly improved however by formulating the action for $\varphi^+$  directly in continuous imaginary frequencies $i\omega$ and subsequently discretizing it there. The free part of the action then is diagonal in imaginary frequencies and the correct (temperature independent) noninteracting $G^{++}$ is obtained trivially when computed via (still real-valued) Langevin stochastic quantization. No boundary conditions in $i\omega$ need to be specified in the free case. 

Now in the presence of interactions, the $\varphi^4$ term is not diagonal in imaginary frequencies but is diagonal in imaginary time. Therefore we decide to implement the Langevin update in two steps, first computing the drift from the kinetic term of the action in imaginary frequencies, where it is local, then transforming the fields via discrete Fourier transform to non-compact imaginary time, where $\varphi^3$ is evaluated locally before transforming back and evaluating the two drift contributions together as in 
\begin{align}
\partial_s \varphi^+(i\omega,s)=-\frac{\delta S_E^0[\varphi^+]}{\delta \varphi^+(i\omega)}-\int d\tau^\prime \frac{\delta S_E^{\rm int}[\varphi^+]}{\delta \varphi^+(\tau^\prime)} \frac{\delta \varphi^+(\tau^\prime)}{\varphi^+(i\omega)}+\eta^\prime(i\omega,s).
\end{align}
$\langle \eta^\prime(i\omega,s)\rangle=0$ with $\langle \eta^\prime(i\omega)\eta^\prime(\omega^\prime)\rangle=2\delta(\omega+\omega^\prime)$ denotes Gaussian noise.
While no explicit boundary conditions need to be specified in this mixed representation setting, the discrete Fourier transform of course pre-supposes periodic boundary conditions. This is why there still remains a residual finite volume artifact, which however is significantly smaller as compared to computing both the kinetic and interacting drift term directly in imaginary time. 

We still have to answer how information about temperature is relayed from $\varphi_E$ to $\varphi^+$. Since we know that in the free theory $G^{++}$ is temperature independent we cannot simply replace $\varphi^+(t=0)=\varphi_E(\bar{\tau}=0)$ at each step from the concurrent update of $\varphi_E$. Information about temperature only enters into $G^{++}$ in the interacting theory, which is why we implement the above mentioned replacement of $\varphi^+$ when evaluating the interacting drift term. This can be seen as an additional constraint on the drift, which is provided by the finite temperature initial density matrix.

\section{Numerical results}\label{sec-2}

As a first benchmark we check how well spectral properties of the theory are reproduced within the {\it standard compact Euclidean simulation} approach using $\lambda=24$, $m=1$, $\beta=1$ on $N_\tau=16$ grid points.  We collect the values of $G_E(\bar\tau)=\langle \varphi_E(\bar\tau)\varphi_E(0)\rangle$ after each $\Delta s=1$ and accumulate $N_{\rm conf}=10^6$ samples, which leads to statistical errors of around $\Delta G/G\sim10^{-3}$. At the same time we carry out a quantum mechanical (QM) computation where the Schr\"odinger equation for the Euclidean $G_E(\bar\tau)$, as well as the real-time field correlator is solved in the truncated Hilbert-space of $32$ energy eigenstates of the harmonic oscillator \cite{Berges:2006xc}.

\begin{wrapfigure}{r}{0.44\textwidth}
  \begin{center}\vspace{-0.3cm}
\includegraphics[scale=0.5]{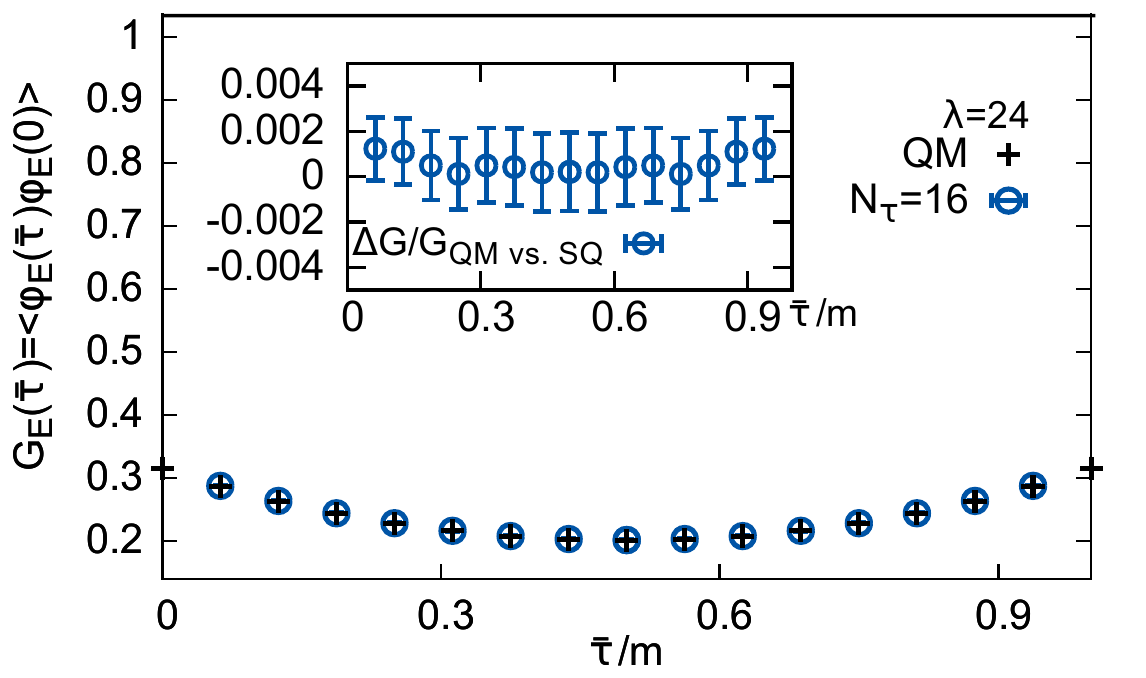}\\
\includegraphics[scale=0.6]{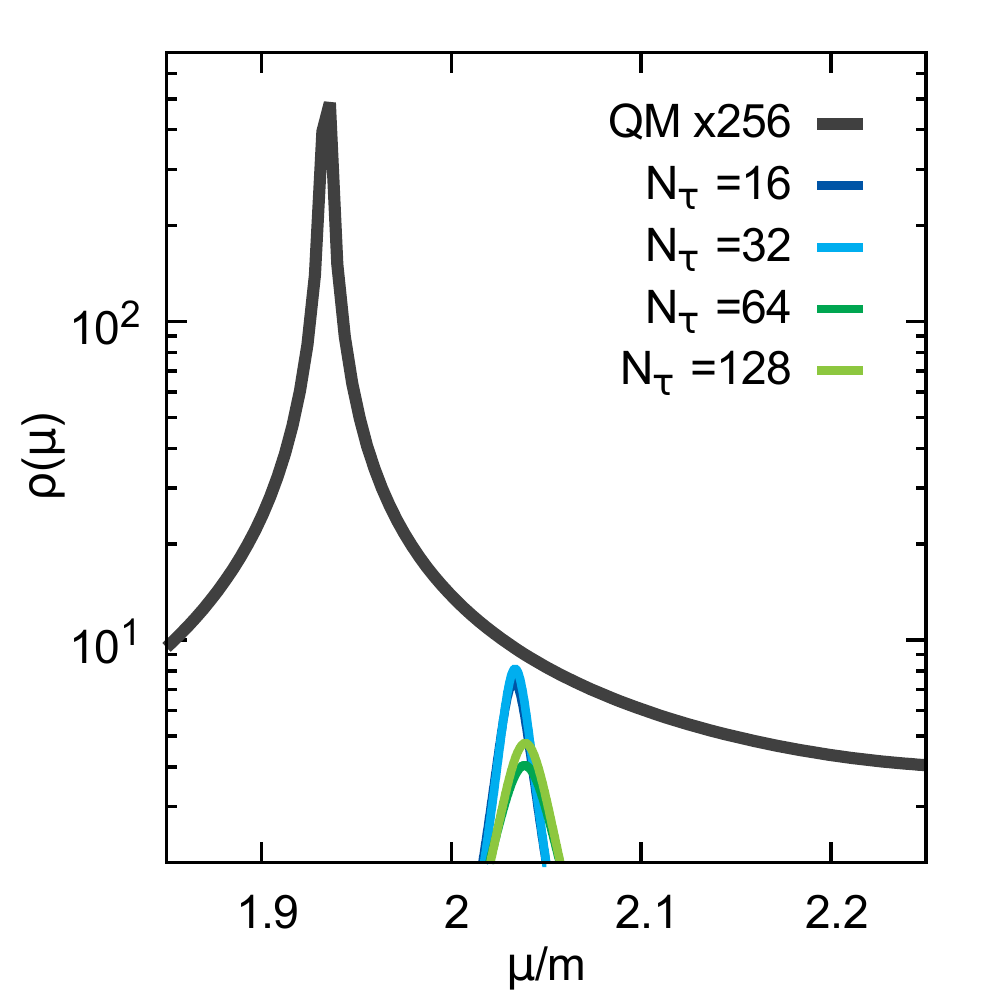}
  \end{center}\vspace{-0.5cm}
 \caption{(top) Comparison of the QM correlator (black cross) to its standard Euclidean simulation (blue circle). (bottom) Reconstructed spectra from Euclidean data via the BR method (colored solid).}\label{Fig:Eucl}\vspace{-0.7cm}
\end{wrapfigure}In Fig.\ref{Fig:Eucl} we show in the top panel the correlator values from that quantum mechanical computation as black crosses. The simulated values from a standard Euclidean stochastic quantization are given by blue circles and the inset confirms that they agree within statistical errors. On the bottom, the QM spectral function from Fourier transforming the numerical solution of the real-time Schr\"odinger equation up to $t/m=1600$ is shown as thick gray curve. The residual width due to the numerical Fourier transform is not relevant here, we focus on the peak position. 

To extract the spectral functions from simulated correlators we deploy the BR method \cite{Burnier:2013nla} with real-time frequencies, denoted here as $\mu\in[10^{-3},200]$ discretized along $N_\mu=4000$ points. Around the low lying peak structure a high resolution window with $N_\mu^{\rm  HP}=1000$ points is selected. The default model we use corresponds to the minimum bias choice of $m(\mu)=1$. For the Euclidean correlator we deploy the naive discretization $K(\mu,\bar\tau,T)={\rm cosh}[\mu(\bar\tau-/2T)]/{\rm  sinh}[\mu/2T]$ of the kernel introduced in eq.\eqref{eq:CorrSpecRel}. The spectrum reconstructed in this way is shown in Fig.\ref{Fig:Eucl} as the dark blue curve in the bottom panel (almost covered by the light blue curve). It is cleat that with relative errors of order $10^{-3}$ and $N_\tau=16$ the peak position is not yet well reproduced. 

In standard simulations, since temperature fixes the Euclidean time extent, the only handle for improvement we have is to go to higher resolution in compact imaginary time, discretizing $\beta=1$ with e.g. $N_\tau=32,64$ or $128$ points. As we argued before this only gives access to higher lying Matsubara frequencies, which are not very sensitive to the physics of the lowest lying spectral peak. And indeed we find that as shown by the three lighter colored curves in the bottom panel of Fig.\ref{Fig:Eucl}, going up to $N_\tau=128$ does not improve the situation. On the other hand, a higher resolution in $\bar{\tau}$ means that the drift term in the Langevin update becomes larger and we have to reduce the Monte-Carlo step size $\Delta s$ significantly to keep the numerics stable, making these finer computation more expensive with limited practical use. 

\begin{figure}[t]
\centering
\includegraphics[scale=0.55]{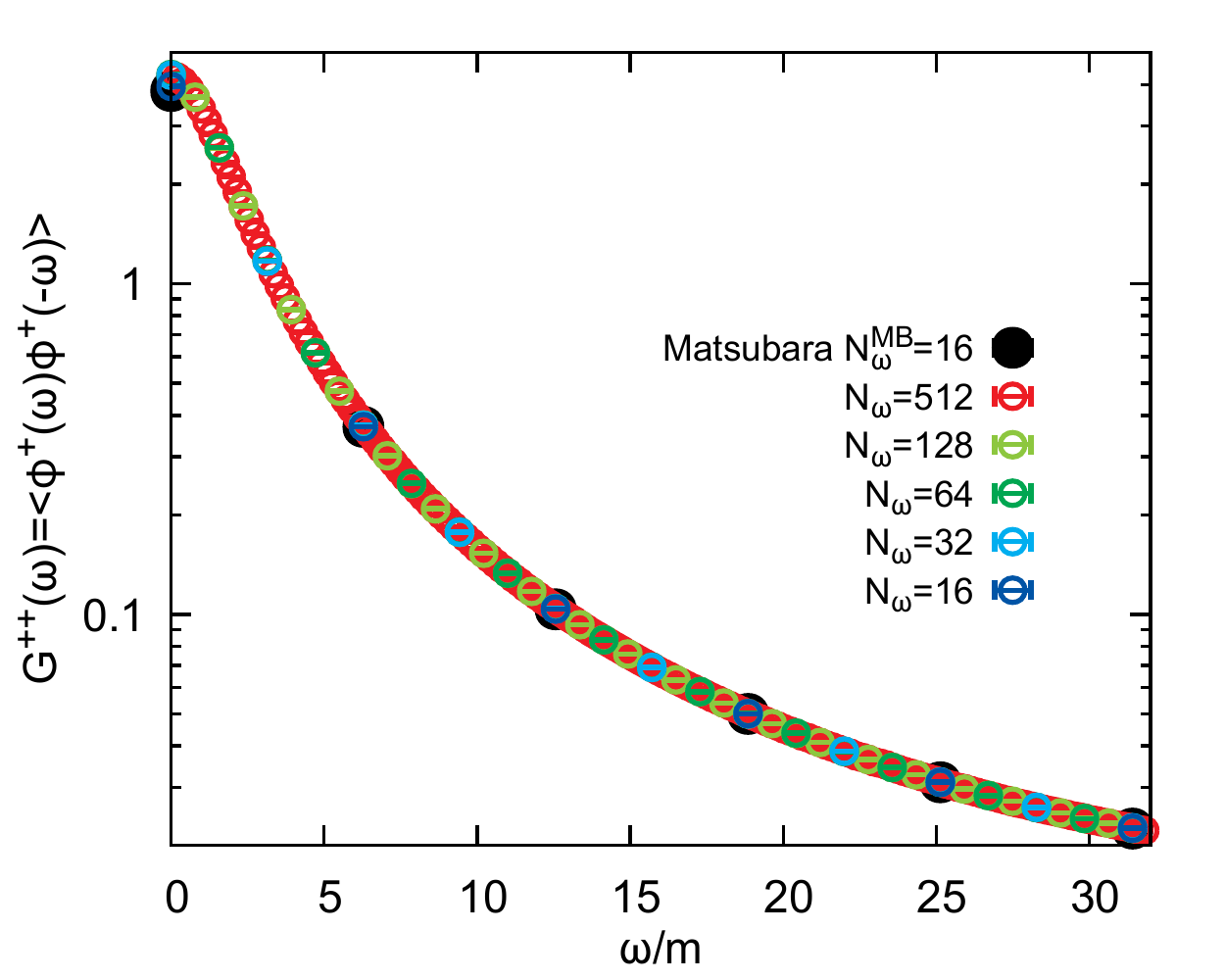}\includegraphics[scale=0.55]{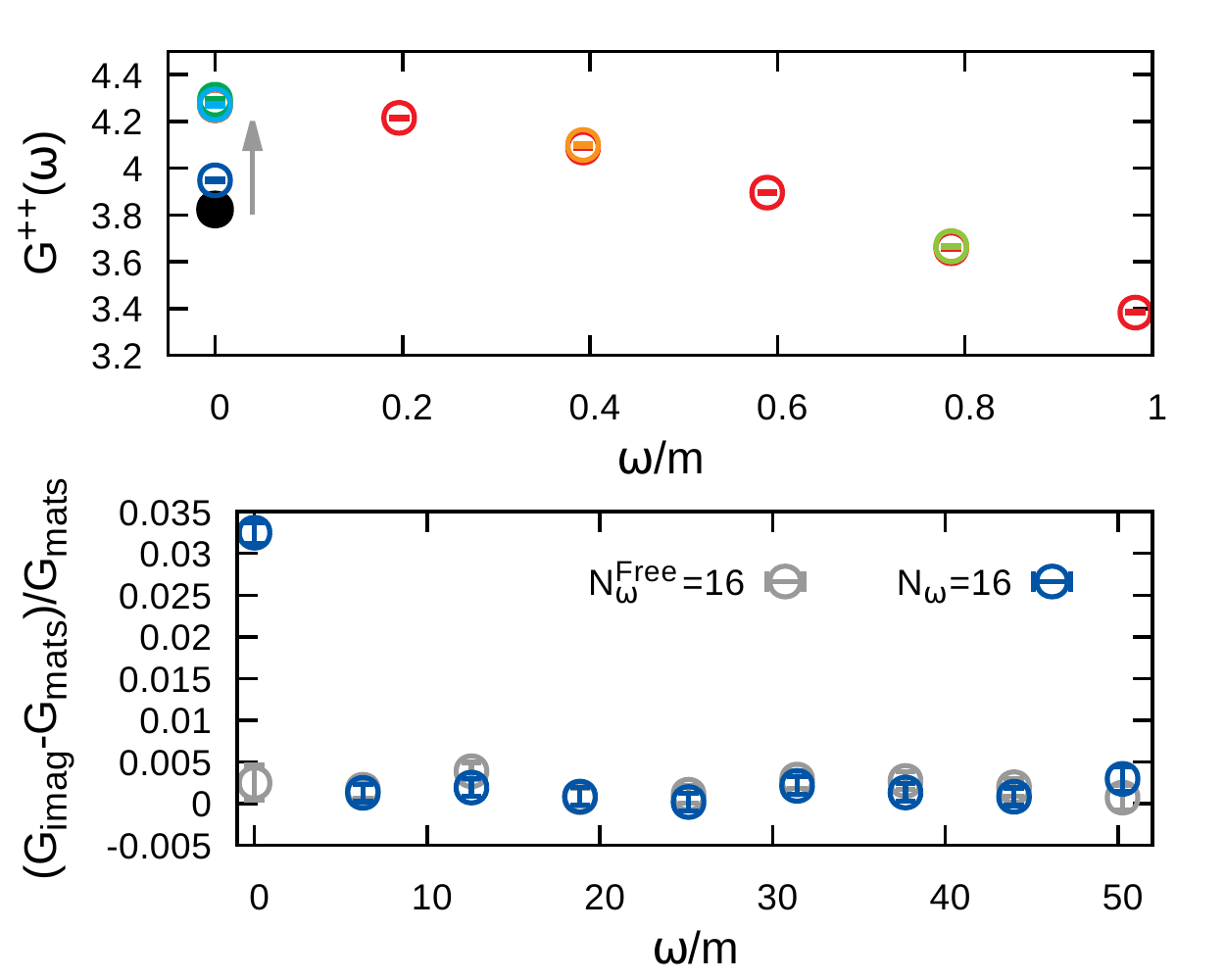}\caption{(left) Values for the Matsubara correlator from standard Euclidean simulations via DFT (black circles) and for $G^{++}$ from the novel simulations in mixed representation (open circles). (right bottom) Difference between the standard Matsubara correlator  (gray) and the novel simulations (blue) at $N_\tau=N_\omega=16$. (right top) Small imaginary frequency behavior of $G^{++}$ in the novel simulations. Note that for increasing extent of the non-compact imaginary time domain, the values converge slightly above the zero Matsubara correlator, a change that turns out to be important to obtain the correct spectral reconstruction.}\label{Fig:FullSims}\vspace{-0.65cm}
\end{figure}

We now turn to the novel simulations, where besides a standard compact Euclidean domain with $N_\tau=16$, we also consider the fields $\varphi^+$ in imaginary frequencies resolved on $N_\omega\geq N_\tau$ points. The kinetic term of the discretized action reads 
\begin{equation}\label{eq:S0}
  S^0=\sum_l \Delta\omega\; \frac{1}{2}\big({\cal P}(\omega_l) + m^2\big)
  |\varphi^+(\omega_l)|^2\,,  
\end{equation}
with the dispersion ${\cal P}(\omega_l)=(2-2{\rm cos}[2\pi l /N_\omega ])/\Delta \bar \tau^2$, $l\in [0,N_\omega-1]$. This choice leads to the same lattice spacing being used in the compact and non-compact imaginary time domain and allows us to compare correlators coming from either the direct imaginary frequency simulations or the DFT of the standard Euclidean correlator. Remember that temperature information enters the $\varphi^+$ update by replacing $\varphi^+(t=0)=\varphi_E(\bar{\tau}=0)$ in the evaluation of the drift.

As is shown in the bottom left panel of Fig.\ref{Fig:FullSims}, resolving $G^{++}$ at the same $N_\omega=16$ imaginary frequencies (blue) as the Matsubara correlator (gray), leads to values, which agree among the two for all finite $\omega_n>0$ within uncertainties. \begin{wrapfigure}{r}{0.44\textwidth}
  \begin{center}\vspace{-0.8cm}
\includegraphics[scale=0.65]{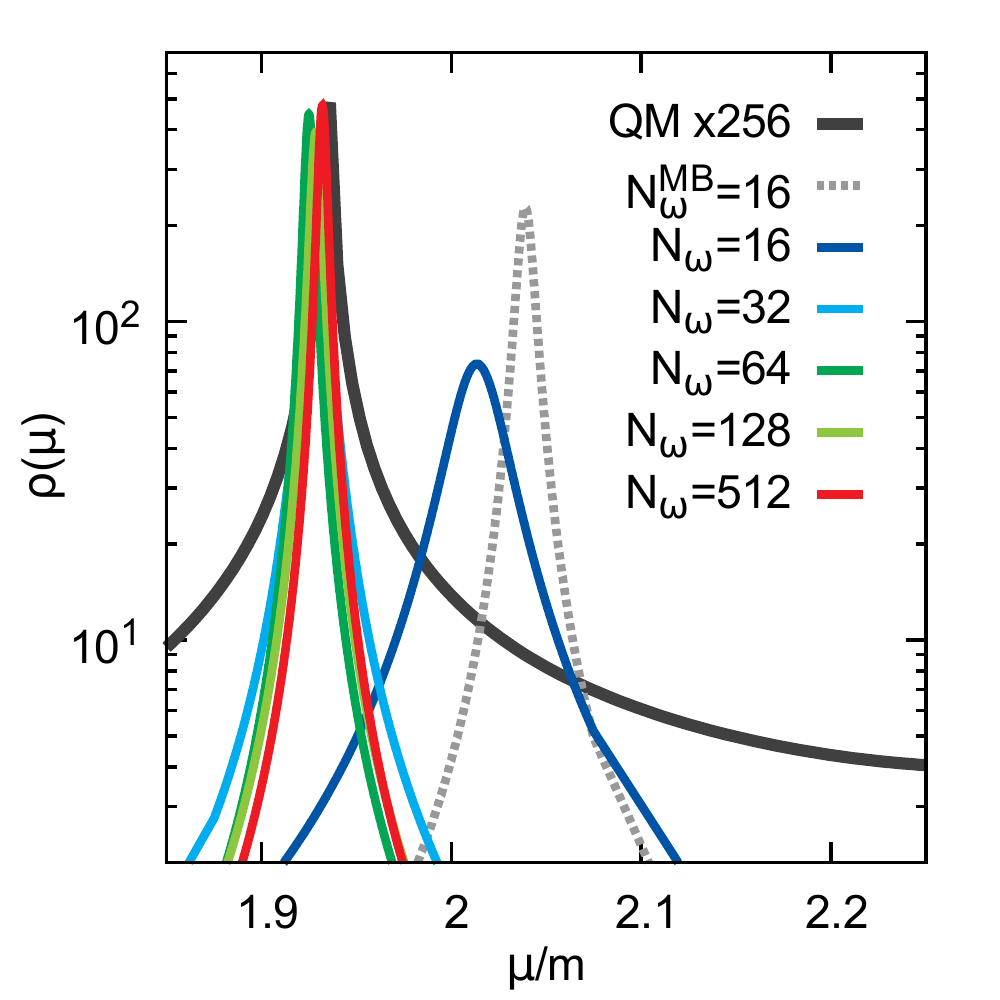}  \end{center}\vspace{-0.5cm}
 \caption{BR method spectral reconstruction from standard Matsubara correlators (gray) and from the novel simulations (colored solid). While access only to Matsubara frequencies (gray, Euclidean) and (blue, new approach) does not yield satisfactory results, resolving between the $\omega_n$'s dramatically improves the reconstruction success.}\label{Fig:FullRecs}\vspace{-1.1cm}
\end{wrapfigure} Interestingly, close to $i\omega=0$ $G^{++}$ shows a deviation, which, if $N_\omega$ is increased beyond $32$, settles down to an asymptotic value slightly above the Matsubara correlator. This difference turns out to be an important ingredient to a successful spectral reconstruction. 

For the imaginary frequency spectral reconstruction in the following $K(\mu,\omega)=\frac{2\mu}{\mu^2+{\cal P}(\omega)}$ is used, which corresponds to the lattice analogue of the continuum kernel and leads to better convergence of the reconstruction algorithm than using the naive discretization of eq.\eqref{eq:CorrSpecRel}. Fig.\ref{Fig:FullRecs} again shows the QM spectrum as thick gray line, and the result from the BR method reconstruction, based on the standard Matsubara correlator, as dashed light gray line. We see that similar to the result in the bottom panel of Fig.\ref{Fig:Eucl}, the peak position is not satisfactorily captured with $N_\tau=16$ Matsubara frequencies. Since $G^{++}(i\omega)$ for $N_\omega=16$ agrees with its Matsubara counterpart for all finite $\omega_n$ the corresponding spectral reconstruction (blue solid) not surprisingly also misses the peak position. Interestingly, the small deviation at $\omega_0$ has already nudged the peak in the direction of the correct result. 

The strength of our novel simulation approach lies in its capability to resolve between the conventional Matsubara frequencies. As shown in the left panel of Fig.\ref{Fig:FullSims} going from $N_\omega=16$ to $N_\omega=512$ eventually produces a collection of points, which appear to smoothly cover the whole imaginary frequency range.  From the discussion of eq.\eqref{eq:CorrSpecRel} we expect that access to the correlator in between the first and second Matsubara frequency will significantly improve the spectral reconstruction success. And indeed as is shown by the colored curves in Fig.\ref{Fig:FullRecs}, for this simulation parameter set,  already twice the standard imaginary frequency resolution $N_\omega=32$ moves the peak much closer to the correct position. Going to even larger numbers, such as $N_\omega=512$ allows us to reproduce the peak position with a deviation of only around one per mille.

\begin{figure}[b]
\center
\includegraphics[scale=0.63]{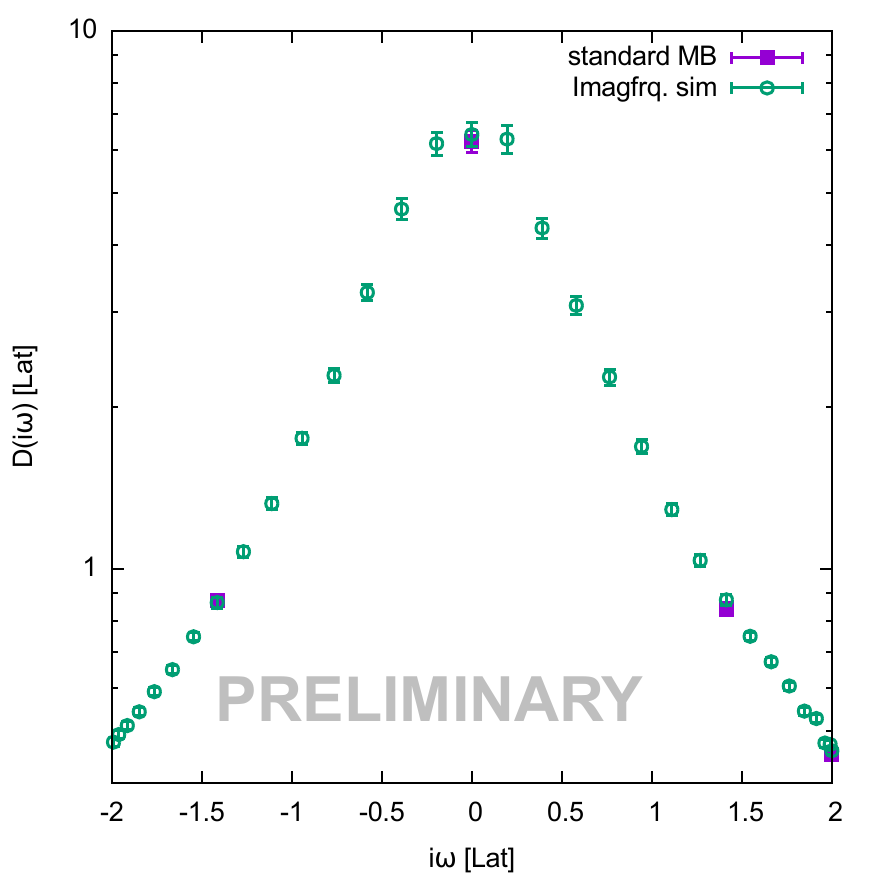}
\includegraphics[scale=0.63]{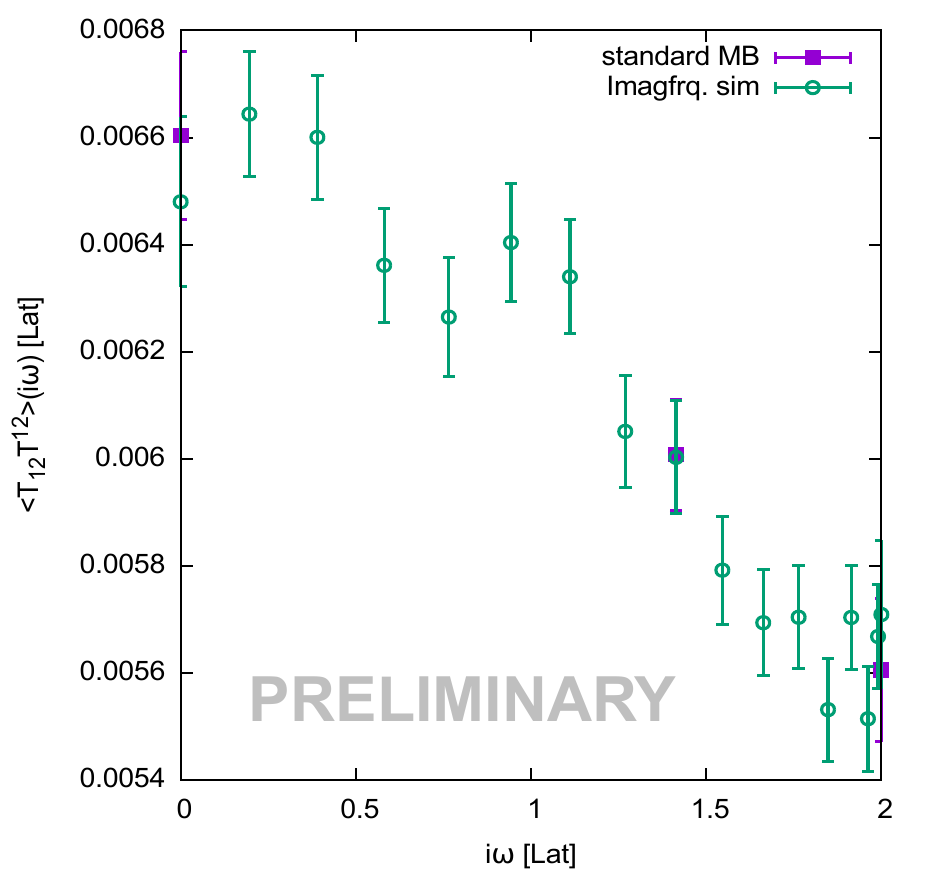}
\caption{Preliminary results for the (3+1)d complex scalar field in imaginary frequencies. We show (left) the field correlator $D(i\omega, {\mathbf p}=0)$, as well as (right) the energy momentum tensor correlator relevant for shear viscosity. Note that the latter is not affected by an exponential decay of the signal to noise ratio but still shows significantly larger errors than the simple field two-point function.}\label{Fig:ExploreScalar}
\end{figure}

While these results are encouraging they have been obtained in a simple toy model only. Therefore it is paramount to extend the computations performed here to genuine field theories. We are currently working on the straight forward implementation of code both for real-valued scalar fields in (2+1) dimensions, as well as for complex scalar fields in (3+1) dimensions. While the focus in the former case lies in investigating bound state spectra, the latter will provide a useful and realistic testing ground for the computation of transport properties. Indeed for the charged scalar field the relevant energy momentum tensor correlator for the shear viscosity \cite{Caracciolo:1988hc} can be written as 
\begin{align}
\langle T_{12}T^{12}\rangle\approx\langle (T_{11}-T_{22}) (T^{11}-T^{22})\rangle \quad {\rm with} \quad
 (T_{11}-T_{22})=\partial_1\varphi(x)\partial_1\varphi^*(x)-\partial_2\varphi(x)\partial_2\varphi^*(x).
\end{align}
First preliminary results for a lightweight and purely exploratory setup of $\lambda=1$, $m^2=0.2$ on $8^3\times N_\tau\times N_\omega$ grids with $N_\tau=4$ and $N_\omega=32$ are shown in Fig.\ref{Fig:ExploreScalar}. On the left we plot the field two-point function $D(p)=\langle \varphi^+(p)\varphi^+(p)\rangle$ at vanishing spatial momentum. The outcome of the standard Euclidean simulation is denoted by the filled squares, while the novel imaginary frequency simulation produces the open circles.  The correlator is not exactly symmetric, since the real- and imaginary part of the scalar field contribute independently to the positive and negative imaginary frequency sides. At least in the symmetric phase, in which the simulation is situated, symmetry emerges better and better with increased statistics. Interestingly, it seems that here the agreement between the different simulations is better, in particular we do not find a statistically significant deviation at $\omega_0$. There are two possible explanations for this, either finite temperature effects are not as relevant for the chosen parameter set or due to the still sizable errorbars, we are just not yet able to resolve a difference that is nevertheless present. Both increasing statistics, as well as testing different parameter sets for higher temperatures is currently high priority work in progress.

As last item we take a look at the preliminary result for the energy momentum tensor on the right in Fig.\ref{Fig:ExploreScalar}. This correlator is exactly symmetric, which is why only its positive frequencies are shown. As expected, this observable, while not suffering from an exponentially decaying signal in imaginary frequencies, still shows much more statistical variance than the simple field two-point functions. It is a pleasure to observe that even though the statistical errors are still sizable, also for this correlator we find agreement between the novel and the standard compact Euclidean simulations on the Matsubara frequencies. A more quantitative study of the correlator and a first spectral reconstruction will be attempted, once adequate statistics of the order of $\Delta D/D\sim10^{-2}$ have been achieved.

In summary, we have presented a novel approach to treating thermal fields as an initial value problem in an additional non-compact Euclidean time. We have formulated a simulation prescription based on this picture, in which the fields $\varphi^+$ and $\varphi_E$ are concurrently sampled using a real-valued Langevin update. To mitigate the effects of finite volume artifacts related to the finite extent of the new imaginary time axis, we simulate $\varphi^+$ directly in imaginary frequencies, where the free theory is diagonalized. Temperature information is supplied to the the $\varphi^+$ fields by imprinting the information of $\varphi_E(\bar\tau=0)$ onto the drift term arising from the interaction part of the action in the Langevin update. Our results in a (0+1) dimensional toy model show that resolving the correlator between the conventional Matsubara frequencies indeed dramatically improves the reconstruction of spectral features. It however will be important to further clarify the meaning of the deviations observed between the correlators produced in the novel and standard simulations around vanishing Matsubara frequencies, in particular since they appear to be important for obtaining the correct spectral reconstruction. Preliminary results for complex scalar fields in (3+1) dimensions are also encouraging and their maturation is work in progress. 

This work is supported by the DFG Collaborative Research Centre "SFB 1225 (ISOQUANT)"

\vspace{-0.2cm}
\bibliography{lattice2017}

\end{document}